\documentclass[]{article}
\usepackage{amsmath}
\usepackage{amssymb}
\usepackage{graphicx}
\usepackage[margin=1.0in]{geometry}

\newcommand{\captionfonts}{\small}

\makeatletter  
\long\def\@makecaption#1#2{%
  \vskip\abovecaptionskip
  \sbox\@tempboxa{{\captionfonts #1: #2}}%
  \ifdim \wd\@tempboxa >\hsize
    {\captionfonts #1: #2\par}
  \else
    \hbox to\hsize{\hfil\box\@tempboxa\hfil}%
  \fi
  \vskip\belowcaptionskip}
\makeatother   

\title{Confronting the Kaya Identity with Investment and Capital Stocks}
\author{Eric Kemp-Benedict\\
Stockholm Environment Institute\\
eric.kemp-benedict@sei-international.org}

\begin{document}

\maketitle
\bibliographystyle{plain}

\begin{abstract}
Scaling relations, such as the IPAT equation and the Kaya identity, are useful for quickly gauging the scale of economic, technological, and demographic changes required to reduce environmental impacts and pressures; in the case of the Kaya identity, the environmental pressure is greenhouse gas emissions. However, when considering large-scale economic transformation, as with a shift to a low-carbon economy, the IPAT and Kaya identities and their cousins fail to capture the legacy of existing capital, on the one hand, and the need for new investment, on the other. While detailed models can capture these factors, they do not allow for rapid exploration of widely different alternatives, which is the appeal of the IPAT and Kaya identities. In this paper we present an extended Kaya identity that includes investment and capital stocks. The identity we propose is a sum of terms, rather than a simple scaling relation. Nevertheless, it allows for quick analysis and rapid exploration of a variety of different possible paths toward a low-carbon economy.

\vspace{2em}

\end{abstract}

\section{Introduction}
In 1972 Ehrlich proposed an identity that captured the main drivers of impacts on the environment \cite{ehrlich_bulletin_1972}. Called the ``IPAT equation'' or ``Ehrlich identity'' it proposes that impact, $I$, is equal to the product of population, $P$, affluence, $A$, and technology, $T$. The IPAT equation is a popular method for decomposing environmental impacts, and has spawned several variants \cite{chertow_ipat_2000,waggoner_framework_2002,york_stirpat_2003}. More recently, structural decomposition of changes in input-output matrices provide a detailed basis for the IPAT identity \cite{baiocchi_understanding_2010}. Within the climate community, the closely-related Kaya identity decomposes carbon dioxide (CO$_2$) emissions into a product of factors: population, GDP per capita, energy intensity per GDP, and emissions intensity per unit energy \cite{nakienovi_special_2000}. Combining some of these factors, the Kaya identity is equivalent to expressing total CO$_2$ emissions, $E$, as a product of GDP, conventionally denoted $Y$, and emissions intensity $\epsilon$,
\begin{equation}\label{kaya_compressed}
E = \epsilon Y.
\end{equation}
A typical analysis would examine the factors that enter the Ehrlich or Kaya identities in order to recommend policy or individual action (e.g., \cite{ekins_step_2004,fan_analyzing_2006,jackson_prosperity_2009,raupach_global_2007}).

In this paper we argue that the IPAT and Kaya identities, while useful, are deficient when they are used to evaluate the need for or potential of substantial economic transformations. This is because they do not take into account the legacy of existing capital and the investments required for such a transformation. Popular movements in high-income countries in support of zero economic growth \cite{jackson_prosperity_2009} or de-growth \cite{martinez-alier_sustainable_2010} (or, perhaps more appropriately, ``a-growth'', meaning ``indifferent to growth'', as proposed in \cite{van_den_bergh_environment_2011}), sometimes assume a return to low-technology modes of consumption and production, but many are premised on the development and dissemination of new technologies, and the retirement of older technologies \cite{chertow_ipat_2000,iea_energy_2010,iea_world_2010,vollebergh_role_2005}. The Kaya identity hides the changes in investment flows and capital stocks required for such a transformation.

To illustrate the difficulties, suppose that a high-income country introduces policies designed to reduce emissions by 80 per cent relative to 2010 levels by 2050. As part of the policy package, the country adopts a ``Factor 10 by 2050'' strategy whereby emissions intensities in 2050 are one-tenth those in 2010.\footnote{Reductions in material intensity per unit service (MIPS) to one-tenth of present values are promoted by the Factor 10 Institute (http://www.factor10-institute.org). The concept is explained in \cite{hinterberger_dematerialization_1999} and put in the context of other approaches to strategic sustainable development in \cite{robert_strategic_2002}. In this paper we are concerned with carbon emissions per unit economic activity, which are not the same as MIPS. Most important, carbon emissions face no fundamental physical constraints and net emissions can even be negative. We use it here because it is a prominent, and physically achievable, benchmark value for thinking broadly about alternative futures.} The Kaya identity suggests that to meet these goals the economy must grow at or below 1.7 per cent per year. While this is a mathematical necessity given the assumptions, it is not clear whether such substantial emissions intensity reductions are consistent with that rate of growth.

In this paper we propose an alternative to the Kaya identity that includes consumption, capital stocks, and investment. It must of necessity include a sum of more than one term. Thus, unlike the IPAT and Kaya identities, which are simple scaling relationships, the formulation that we propose involves relative changes in different terms, and is therefore more difficult to use than IPAT-like expressions. We argue that the additional complexity is unavoidable, because, as we demonstrate, it captures an interrelated set of structural changes in the economy.

\section{CO$_2$ emissions from economic activities}
We express total CO$_2$ emissions $E(t)$ as a sum of three terms: one for emissions from operating capital, $E_K(t)$, such as industrial machinery, power stations, and some transport infrastructure; one for immediate consumption, $E_C(t)$; and one for investment, $E_I(t)$,
\begin{subequations}\label{total_emissions}
\begin{align}
E(t) &= E_K(t) + E_C(t) + E_I(t)\label{total_emissions_a}\\
&= E_K(t) + \epsilon_C(t) C(t) + \epsilon_I(t) I(t)\label{total_emissions_b}\\
&= Y(t) \left[\bar{\epsilon}_K(t) + \left(1-s(t)\right)\epsilon_C(t) + s(t)\epsilon_I(t) \right]\label{total_emissions_c}.
\end{align}
\end{subequations}
In passing from Equation (\ref{total_emissions_a}) to Equation (\ref{total_emissions_b}) we assume that emissions from consumption and investment activities can be expressed in terms of the volume of current consumption and investment, with time-varying emissions coefficients $\epsilon_C(t)$ and $\epsilon_I(t)$; that is, we treat consumption and investment as flows, with instantaneous emissions associated with those flows. In contrast, as we explain below, emissions from capital are best thought of as a stock that depends on the quality of the capital put in place with prior investment. In passing from Equation (\ref{total_emissions_b}) to Equation (\ref{total_emissions_c}), we assume that total economic output $Y(t)$ is either saved or consumed, and that all savings result in investment. The savings rate $s(t)$ is therefore given by $I(t)/Y(t)$ and, as indicated, it may change over time.

\begin{figure}[h]
\centering
\includegraphics[scale=0.65]{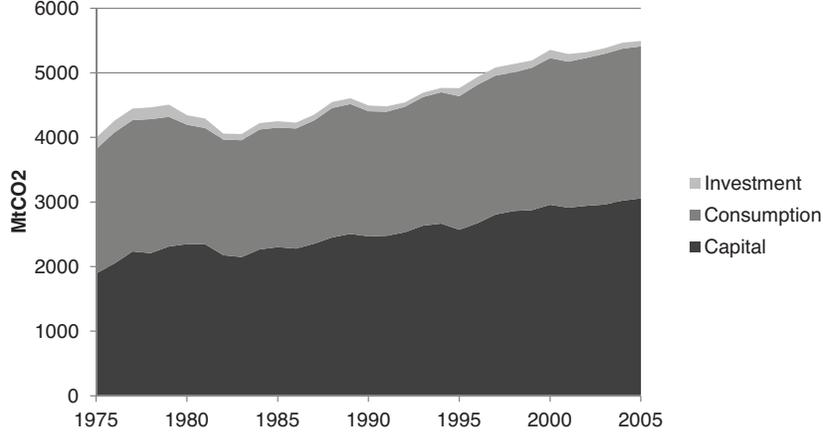}
\caption{Historical emissions from fuel combustion and cement manufacture in the US}
\label{hist_emiss}
\end{figure}

Carbon dioxide emissions from energy combustion by economic sector are available for most countries from the International Energy Agency (IEA) \cite{iea_co2_2007}, while emissions from cement manufacture are available from the Climate Dioxide Information Analysis Center (CDIAC) \cite{boden_global_2010}. As a proxy for emissions from operating capital we calculate total energy combustion emissions from the energy sector, from rail and pipeline transport, and from manufacturing (excluding iron and steel and machinery manufacture). For emissions from investment activities we add emissions from cement manufacture to energy combustion emissions from the iron and steel and machinery manufacture sectors. Finally, for consumption emissions we include energy combustion emissions from air and road transport, households, the service sector, and agriculture. With these proxies, we estimate that of total 2005 emissions in the United States, 58.2 per cent was due to the operation of capital, 40.4 per cent due to consumption, and 1.4 per cent due to investment. The trend in emissions over the 30 years from 1975 to 2005 is shown in Figure \ref{hist_emiss}. We note that while emissions due to investment activities are small, if a low-carbon economy requires significant new investment, and other emissions drop strongly, they may not remain small as a share of the total, as it is very difficult to reduce emissions from iron and steel and cement manufacture.

\subsection{Capital Investment}
Before formulating an expression for emissions from operating capital as a stock, we first consider (physical) capital itself, $K(t)$, which is a standard exemplar of a stock, fed by a flow of investments and drained through depreciation,
\begin{equation}\label{capital_balance}
\frac{dK(t)}{dt} = - \delta K(t) + I(t).
\end{equation}
As indicated in the equation, we assume the the depreciation rate $\delta$ is constant in time.

A key measure of the structure of an economy is its capital intensity, $k(t)$, which is given by the ratio $K(t)/Y(t)$. The capital intensity, which has units of time, captures the extent to which production relies on physical capital, rather than other inputs, notably labour. We therefore divide Equation (\ref{capital_balance}) by the size of the economy, $Y(t)$. This gives
\begin{equation}\label{capital_balance_div_by_Y}
\frac{1}{Y(t)}\frac{dK(t)}{dt} = - \delta k(t) + s(t).
\end{equation}
This equation can be recast as a dynamic equation for the capital intensity $k(t)$. Consider
\begin{subequations}\label{time_change_k_prelim}
\begin{align}
\frac{dk(t)}{dt} &= \frac{d\,}{dt} \frac{K(t)}{Y(t)}\label{time_change_k_prelim_a}\\
&= \frac{1}{Y(t)}\frac{dK(t)}{dt} - k(t) \frac{1}{Y(t)}\frac{dY(t)}{dt}.\label{time_change_k_prelim_b}
\end{align}
\end{subequations}
Note that $(1/Y(t))dY(t)/dt = d/dt \ln Y(t)$ is the instantaneous growth rate of the economy, which we denote by $r(t)$. Substituting into Equation (\ref{capital_balance_div_by_Y}), we find
\begin{equation}\label{time_change_capital_intens}
\frac{dk(t)}{dt} + r(t)k(t) = - \delta k(t) + s(t).
\end{equation}
This equation can be integrated---in the case of constant $r(t)$ and $s(t)$ it can be integrated exactly---but more interesting is the asymptotic value $k_\infty$ of the capital intensity of the economy at constant growth and savings rates, which can be found by setting the derivative of $k(t)$ in Equation (\ref{time_change_capital_intens}) equal to zero and solving for $k$,
\begin{equation}\label{k_infty}
k_\infty = \frac{s}{r+\delta}.
\end{equation}
Because the growth rate and savings rate are assumed constant, we have dropped the time argument in this equation. This equation, a standard result in economic growth theory, is interesting because it relates the long-term capital intensity of the economy---a key indicator of the type of economy---to the savings, growth, and depreciation rates.

\subsection{Emissions from Capital as a Stock}
As new capital is put in place through investment, it has the emissions profile corresponding to the technology at the time the investment is made, aside from the changes in emissions that arise from deterioration and upgrading. For simplicity, we ignore deterioration, and consider upgrades to old equipment as new investment. We can therefore treat the emissions from operating capital as something that has been put in place through historical flows of investment and removal---that is, the stock of capital corresponds to a stock of emissions from operating that capital. If new capital has an emissions intensity $\epsilon_K(t)$, then
\begin{equation}\label{capital_emiss_balance}
\frac{dE_K(t)}{dt} = - \delta E_K(t) + \epsilon_K(t) I(t).
\end{equation}
We note that the units of the emissions intensity of new capital $\epsilon_K(t)$ that appears in this equation are the same as those of average emissions from operating capital $\bar{\epsilon}_K(t)$ in Equation (\ref{total_emissions_c}); that is, tonnes of CO$_2$ per USD. We also note that in this equation we assume that the depreciation rate $\delta$ reflects the rate of removal of old capital from the economy. In fact, economic depreciation also captures the effect of wear and tear. While a more refined approach could distinguish these two phenomena, we equate them for simplicity.

From this point the analysis for average emissions from capital, $\bar{\epsilon}_K(t)$, is the same as for capital intensity. Dividing Equation (\ref{capital_emiss_balance}) by the size of the economy $Y(t)$ gives
\begin{equation}\label{capital_emiss_balance_per_Y}
\frac{1}{Y(t)}\frac{dE_K(t)}{dt} = - \delta\bar{\epsilon}_K(t) + \epsilon_K(t) s(t),
\end{equation}
where $\bar{\epsilon}_K(t)$ is emissions from capital divided by the size of the economy, as in Equation (\ref{total_emissions_c}). By analogy with Equations (\ref{time_change_k_prelim}), this can be shown to be equivalent to
\begin{equation}\label{time_change_epsilon_K}
\frac{d\bar{\epsilon}_K(t)}{dt} + r(t) \bar{\epsilon}_K(t) = - \delta\bar{\epsilon}_K(t) + \epsilon_K(t) s(t).
\end{equation}
This equation can be integrated if the time rate of change of $\epsilon_K(t)$, $r(t)$, and $s(t)$ are known. We assume that the emissions intensity of new capital declines exponentially at a rate $\alpha$; that is,
\begin{equation}\label{exp_decline_alpha}
\epsilon_K(t) = \epsilon_{K0}e^{-\alpha t}.
\end{equation}
We take historical data for the growth rate $r(t)$ and savings rate $s(t)$ when fitting the model, and assume they are constant in time when exploring possible futures. We also assume that $\alpha$ is constant in time; while we do not calculate formal statistics when fitting our model, the historical data we have compiled for the US appears consistent with an assumption of stable parameters. Using standard techniques, the solution to Equation (\ref{time_change_epsilon_K}) is
\begin{equation}\label{epsilon_k_full_equation}
\bar{\epsilon}_K(t) = e^{-\delta t - \int_0^t dt'\,r(t')}\left[
   \bar{\epsilon}_{K0} + \epsilon_{K0} \int\limits_{0}^{t}dt'\,e^{-(\alpha + \delta) t' - \int_0^{t'} dt''\,r(t'')} s(t')
\right].
\end{equation}
When applied to historical data we evaluate the integrals numerically (using Newton's method). In the case of constant $r$ and $s$, this expression can be evaluated exactly, to give
\begin{equation}\label{epsilon_K}
\bar{\epsilon}_K(t) = \left(\bar{\epsilon}_{K0} + s\epsilon_{K0}f(t)\right)e^{-(r+\delta)t},
\end{equation}
where
\begin{equation}\label{def_of_ft}
f(t) = \left\{
\begin{matrix}
t &\mbox{ if $r+\delta-\alpha=0$} \\
\frac{1}{r+\delta-\alpha}\left[e^{(r+\delta-\alpha)t}-1\right] &\mbox{ otherwise}
\end{matrix} \right. .
\end{equation}

\begin{figure}[h]
\centering
\includegraphics[scale=0.65]{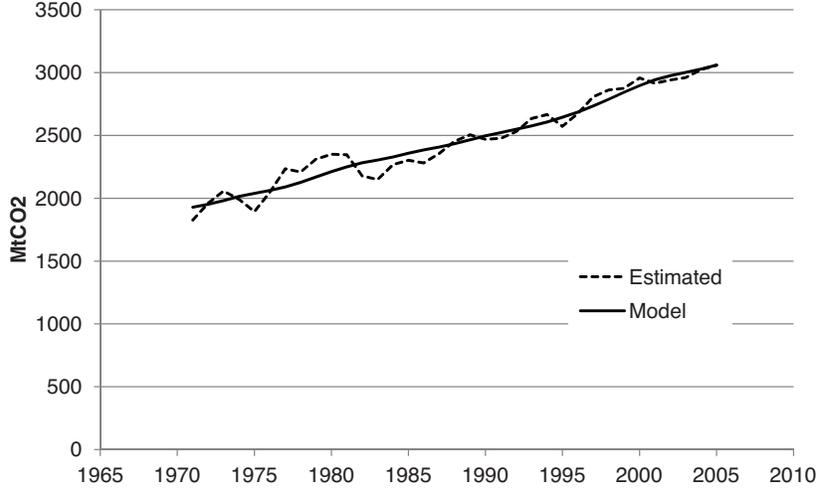}
\caption{Modelled and estimated emissions from operating capital in the US}
\label{model_vs_observed_capital}
\end{figure}

We fit Equation (\ref{epsilon_k_full_equation}) using historical data from the US. As for Figure \ref{hist_emiss}, we used data on emissions from combustion from IEA \cite{iea_co2_2007} and from cement manufacture from CDIAC \cite{boden_global_2010}. For the savings rate and GDP growth rate we used data from the World Bank \cite{world_bank_world_2011}. We constructed a time series from 1971, when IEA began reporting disaggregated emissions by manufacturing sector, through 2005. For later convenience when generating future scenarios, we set the reference time $t=0$ to the final year of the time series, 2005, rather than the initial year, 1971. Minimizing the mean squared deviation using Microsoft Excel 2007's Solver utility, we estimated $\alpha = 0.9$ per cent per year, $\delta = 3.7$ per cent per year (corresponding to a time constant $1/\delta = 27$ years), $\bar{\epsilon}_{K0} = 3059$ MtCO$_2$, and $\epsilon_{K0} = 1029$ MtCO$_2$ per year. The fitted model and historical data are shown in Figure \ref{model_vs_observed_capital}.

\subsection{Total Emissions}
Referring to Equation (\ref{total_emissions_c}), and substituting for $\bar{\epsilon}_K(t)$ from Equation (\ref{epsilon_K}), gives
\begin{equation}\label{total_emissions_ver1}
E(t) = Y(t) \left[\left(\bar{\epsilon}_{K0} + s\epsilon_{K0}f(t)\right)e^{-(r+\delta)t} + \left(1-s\right)\epsilon_C(t) + s\epsilon_I(t) \right],
\end{equation}
where we assume a constant savings rate $s$, as in Equation (\ref{epsilon_K}). Assuming that the emissions intensity of consumption declines exponentially at a rate $\beta$ and emissions from investment at a rate $\gamma$, this can be written
\begin{equation}\label{total_emissions_ver2}
E = Y(t) \left[\left(\bar{\epsilon}_{K0} + s\epsilon_{K0}f(t)\right)e^{-(r+\delta)t} + (1-s)\epsilon_{C0}e^{-\beta t} + s\epsilon_{I0}e^{-\gamma t} \right].
\end{equation}
Fitting historical data for log consumption emissions from 1971 to 2005 against time using ordinary least squares yields an estimated value for $\beta$ of 2.8 per cent per year. The historical coefficient of investment emissions, $\epsilon_{I}(t)$, shows two distinct periods: a rapid decline from 1971 to 1981, and a much slower decline since. Fitting log investment emissions against time using data from the period 1982 to 2005 gives an estimated value for $\gamma$ of 2.0 per cent per year.

\section{Alternative Futures}
We now apply the model developed in this paper to the future of the US economy to 2050 under different assumptions for rates of decline in emissions from consumption and new capital ($\alpha$, $\beta$), the depreciation rate $\delta$, growth rate $r$, and long-term capital intensity $k_\infty$ relative to the present. From Equation (\ref{k_infty}) this allows us to calculate an implied savings rate $s$. We assume that the rate of reduction in emissions intensity from investment activities continues as in the past, reflecting the challenges to reducing emissions in the steel and cement sectors.

We present different combinations of parameters descriptively in Table \ref{experiments_text}, and quantitatively in Table \ref{experiments}.

\begin{table}[h!tb]
\centering
\begin{small}
\begin{tabular}{l c c c c c c c }
\hline
 & $r$ & $\delta$ & $k_\infty$ & $\alpha$, $\beta$ & $\gamma$ \\
\hline
Extrapolation & historical & historical & present & historical & historical \\
Accelerated reductions & - & - & - & Factor 10 by 2050 & - \\
Aggressive reductions & - & - & - & Factor 10 by 2030 & - \\
Accelerated retirement & - & accelerated & - & Factor 10 by 2030 & - \\
Steady state, high tech & zero & accelerated & 50\% higher & Factor 10 by 2030 & - \\
\hline
\end{tabular}
\end{small}
\caption{Descriptions of parameter assumptions for alternative future CO$_2$ emissions}
\label{experiments_text}
\end{table}

\begin{table}[h!tb]
\centering
\begin{small}
\begin{tabular}{l r r r r r r r }
\hline
 & $r$ & $\delta$ & $k_\infty$ & $\alpha$ & $\beta$ & $\gamma$ & $s$ \\
 & (\%/yr) & (\%/yr) & & (\%/yr) & (\%/yr) & (\%/yr) & (\%, calc.) \\
\hline
Extrapolation & 3.1 & 3.7 & 1.0 & 0.9 & 2.8 & 2.0 & 14 \\
Accelerated reductions & 3.1 & 3.7 & 1.0 & 5.6 & 5.6 & 2.0 & 14 \\
Aggressive reductions & 3.1 & 3.7 & 1.0 & 10.9 & 10.9 & 2.0 & 14 \\
Accelerated retirement & 3.1 & 10.0 & 1.0 & 10.9 & 10.9 & 2.0 & 27 \\
Steady state, high tech & 0.0 & 10.0 & 1.5 & 10.9 & 10.9 & 2.0 & 31 \\
\hline
\end{tabular}
\end{small}
\caption{Parameter sets for alternative future CO$_2$ emissions}
\label{experiments}
\end{table}

\begin{figure}[h!tb]
\centering
\includegraphics[scale=0.65]{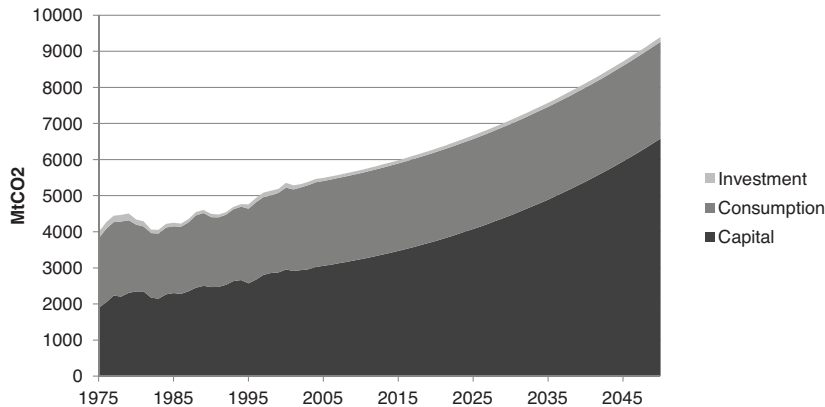}
\caption{Extrapolation: historical and future emissions assuming no change in parameters}
\label{future_exrapolated}
\end{figure}

First, assuming no change in parameter values from 2010 onward (the \emph{extrapolation} set from Table \ref{experiments}) yields continuously growing emissions as shown in Figure \ref{future_exrapolated}. By 2050 in this scenarios emissions for the US are just under 10 Gt CO$_2$ per year. To put this value in context, it is close to the estimated \emph{global} emissions in 2050 consistent with a low (6 to 32 per cent) probability of exceeding 2$^{\circ}$C \cite{meinshausen_greenhouse-gas_2009}, so it is inconsistent with the goal of avoiding dangerous climate change.

\begin{figure}[h!tb]
\centering
\includegraphics[scale=0.65]{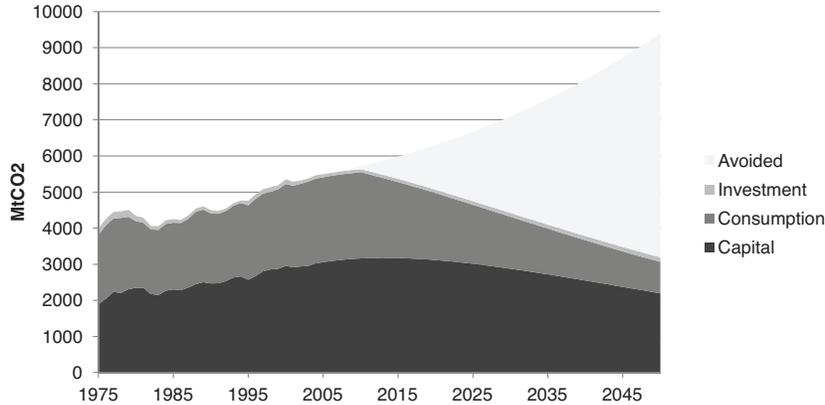}
\caption{Accelerated reductions: historical and future emissions with a ``Factor 10 by 2050'' strategy for new capital and consumption}
\label{future_accel_emiss_reduction}
\end{figure}

Next, suppose that rates of emissions intensity reductions from consumption and new capital follow a ``Factor 10 by 2050'' strategy, as in the parameter set \emph{accelerated reductions}. As explained above, we assume that emissions improvements for investment are difficult to change, and leave the rate at its historical value. This gives the trajectory shown in Figure \ref{future_accel_emiss_reduction}. As seen in the figure, total emissions decline steadily, reaching 71 per cent of their 1990 value in 2050. However, they also decline slowly, suggesting a need for more aggressive action.

\begin{figure}[h!tb]
\centering
\includegraphics[scale=0.65]{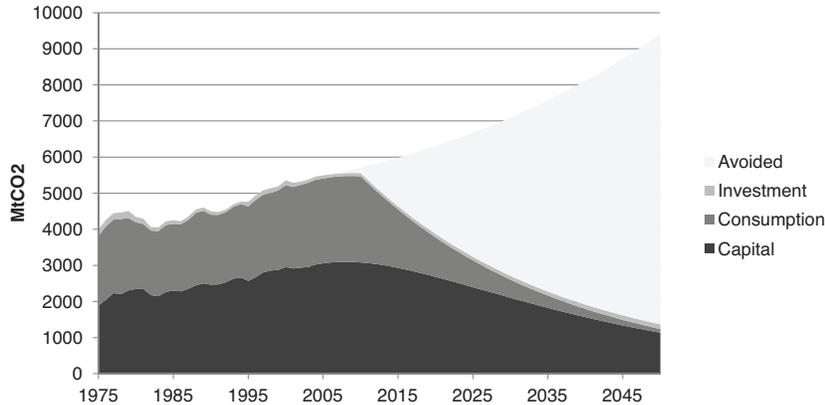}
\caption{Aggressive reductions: historical and future emissions with a ``Factor 10 by 2030'' strategy for new capital and consumption}
\label{future_aggressive_emiss_reduction}
\end{figure}

We consider more rapid reductions in emissions intensity of new capital and consumption in the third parameter set, \emph{aggressive reductions}. In this case we suppose that both $\epsilon_K(t)$ and $\epsilon_C(t)$ decline follow a ``Factor 10 by 2030'' strategy in which they drop to 10 per cent of their 2010 value by 2030 and then a similar drop between 2030 and 2050. The resulting trajectory is shown in Figure \ref{future_aggressive_emiss_reduction}. In this figure emissions from capital and consumption both decline through the period, bringing total emissions in 2050 to 30 per cent of 1990 emissions. This is a considerable reduction, but given a global cap of 10 GtCO$_2$ per year by 2050, it is still not sufficient.

\begin{figure}[h!tb]
\centering
\includegraphics[scale=0.65]{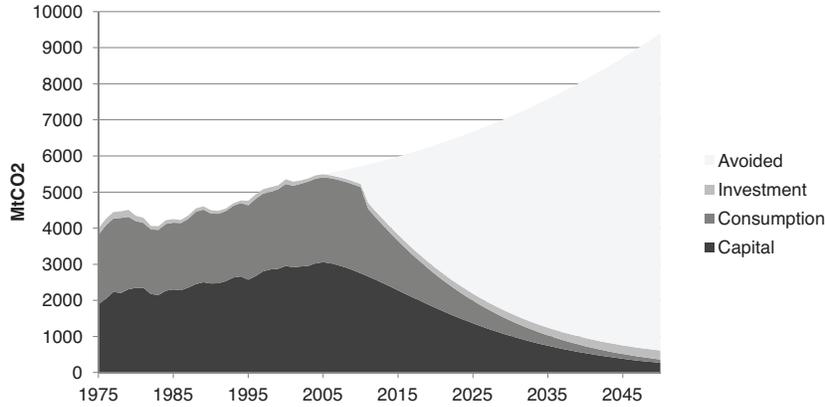}
\caption{Accelerated retirement: historical and future emissions with a ``Factor 10 by 2030'' strategy for new capital and consumption and faster-than-historical capital depreciation}
\label{future_accel_capital_retirement}
\end{figure}

With the measures taken in the \emph{aggressive reductions} parameter set the impact of legacy capital becomes clear. Capital emissions do not fall as quickly as consumption emissions because the emissions profile has already been determined by previous investment. Therefore, in the \emph{accelerated retirement} parameter set we assume that the depreciation rate is increased to 10 per cent per year. Keeping the economic growth rate and long-term capital intensity at their 2010 levels, this implies a substantial---but not unprecedented---increase in the savings rate, from 14 per cent per year to 27 per cent per year. While this is high, rates in rapidly-growing Asian economies have been higher \cite{world_bank_world_2011}. The results are shown in Figure \ref{future_accel_capital_retirement}. As seen in the figure, emissions from capital and consumption both fall more quickly than in the previous cases. Emissions from capital fall more rapidly because older, ``dirtier'' capital is removed from production; emissions from consumption fall more quickly because people reduce their consumption in favour of saving to replace the retired capital. Emissions from investment activities increase, but total emissions in 2050 are only 14 per cent of 1990 levels.

\begin{figure}[h!tb]
\centering
\includegraphics[scale=0.65]{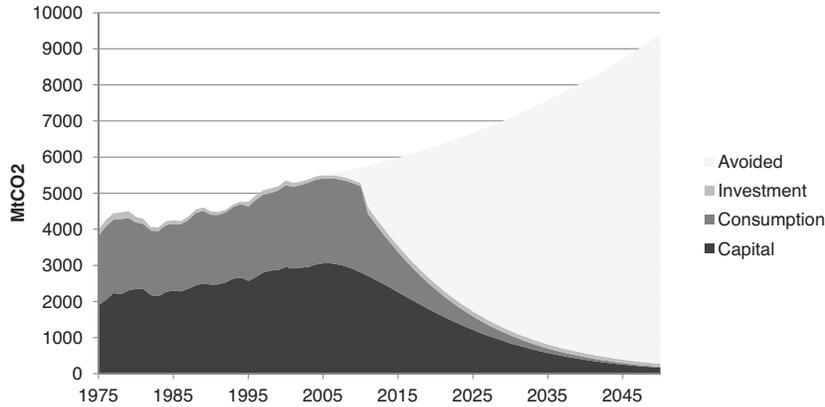}
\caption{Steady state, high tech: historical and future emissions with high rates of emissions intensity reduction, zero growth, and increased capital intensity}
\label{future_steadystate_hightech}
\end{figure}

Going beyond the emissions reductions shown in Figure \ref{future_accel_capital_retirement} can be accomplished in more than one way. We present one option in the \emph{steady state, high tech} parameter set. In this case emissions from new capital, consumption, and investment all decline at higher than historical rates. The economy becomes ``steady-state'' in that the growth rate is zero, but it also becomes more capital intensive, the long-term capital intensity being 50 per cent higher than in the past. The implied savings rate is 31 per cent---high, but still not unprecedented. The results are shown in Figure \ref{future_steadystate_hightech}. As seen in the figure, emissions drop sharply from 2011 to 2050, reaching 6 per cent of 1990 levels by the end of the period. The trajectory in Figure \ref{future_steadystate_hightech} is not very different from that in Figure \ref{future_accel_capital_retirement}, suggesting that strong emissions reductions are not (in principle) incompatible with a growing economy, but as the principle has never been active in the past, there is reason to be sceptical that it will be followed in the future \cite{jackson_prosperity_2009}.

\section{Discussion and Conclusion}
We have extended the Kaya identity to take into account the effects of investment activities and legacy capital on carbon emissions. Unlike the Kaya identity, the identity in this paper is a sum of three terms, representing emissions from the operation of capital, consumption, and investment. It is therefore more difficult to manipulate than the Kaya identity, and less appropriate for ``back-of-the-envelope'' calculations. Nevertheless, we have tried to keep the formulation as close to the spirit and form of the Kaya identity as possible, and the resulting formula is, compared to a full economic model, relatively easy to use. Also, it captures an important reality of the climate challenge, that highly industrialized economies must retire capital early and turn it over rapidly in order to capture the benefits of improved capital equipment.

We applied our identity to the US economy and generated possible future emissions trajectories. Based on that analysis, we argue that the large emissions reductions required to avoid dangerous climate change will require accelerated turnover of capital and rapid improvements in the emissions profiles of consumption activities and capital equipment. To a lesser degree, it may require efforts to reduce emissions from investment activities.

The identity we propose, which is an accounting identity rather than a model, can represent alternatives being discussed today in high-income countries, including no growth and de-growth scenarios, and both low-technology and high-technology routes to reducing emissions. It should also be appropriate in rapidly-growing low-income and middle-income countries that are currently investing in new capital equipment. An accounting identity, whether ours or the Kaya identity, cannot replace a full technical and economic analysis, but can provide a rough estimate of the magnitude of change required to meet a target emissions level. While not conclusive, the reasonable fit shown in Figure \ref{model_vs_observed_capital} suggests enough stability in the estimated parameters (for the US economy, at least) that our proposed identity can be useful for broadly delimiting the requirements for a low-carbon economy.

\bibliography{ipat_with_capital}

\end{document}